\RequirePackage{lineno}
\documentclass[aps,superscriptaddress,showpacs,preprint,amsmath,amssymb]{revtex4}
\usepackage{epsfig}
\usepackage{epsf}
\usepackage{CJK,upgreek,fancyhdr}
\usepackage{bm}        
\usepackage{amssymb}   
\usepackage{amsmath}
\usepackage{array}
\usepackage{multirow}
\usepackage{color}

\newcommand{\PreserveBackslash}[1]{\let\temp=\\#1\let\\=\temp}
\newcolumntype{C}[1]{>{\PreserveBackslash\centering}p{#1}}
\newcolumntype{R}[1]{>{\PreserveBackslash\raggedleft}p{#1}}
\newcolumntype{L}[1]{>{\PreserveBackslash\raggedright}p{#1}}

\def \jp {J/\psi}

\def \ee {e^+e^-}
\def \usig{{\Sigma^{0}}}

\def \half {{1\over 2}}
\def \ee {e^+e^-}
\def \jp {J/\psi}

\def \ldc {{\Lambda_c^+}}
\def \ldcb {{\bar\Lambda_c^-}}
\def \ld {{\Lambda}}
\def \ldb {{\bar\Lambda}}
\def \xic {{\Xi^0_c}}
\def \bc {{\mathcal{B}_c}}

\begin{document}
\begin{CJK*}{GB}{gbsn}
\title{Charmed baryon decay asymmetry in $\ee$ annihilation}
\author{Dan Wang (Íõµ¤)}
\affiliation{University of Chinese Academy of Science, Beijing 100049, China}
\author{Rong-Gang Ping (ƽÈÙ¸Õ)}\email{pingrg@ihep.ac.cn}\affiliation{Institute of High Energy Physics, Chinese Academy of Sciences, Beijing 100049, China}
\author{Lei Li (ÀîÀÙ)}\email{lilei2014@bipt.edu.cn}
\affiliation{Beijing Institute of Petrochemical Technology, Beijing 102617, China}
\author{Xiao-Rui Lyu (ÂÀÏþî£)} \email{xiaorui@ucas.ac.cn}\affiliation{University of Chinese Academy of Science, Beijing 100049, China}\affiliation{CAS Key Laboratory of Vacuum Physics, UCAS, Beijing 100049, China}
\author{Yang-Heng Zheng (Ö£Ñôºã)} \email{zhengyh@ucas.ac.cn}\affiliation{University of Chinese Academy of Science, Beijing 100049, China}\affiliation{CAS Key Laboratory of Vacuum Physics, UCAS, Beijing 100049, China}
\date{\today}
\vspace{1.0cm}
\begin{abstract}
We propose to measure the decay asymmetry parameters in the hadronic weak decays of singly charmed baryons, such as  $\ldc\to\ld\pi^+,\Sigma^0\pi^+,p \bar K_0$, $\Xi_c^0\to\Xi^-\pi^+$ and $\Omega_c^0\to\Omega^-\pi^+$.
The joint angular formulae for these processes are presented, and are used to extract the asymmetry parameters in $\ee$ annihilation data. Base on the current $\ldc$ data set collected at BESIII, we estimate the experimental sensitivities to measure the parameters $\alpha_{\ld\pi^+}$ for $\ldc\to\ld\pi^+$, $\alpha_{\Sigma^+\pi^0}$ for $\ldc\to\Sigma^+\pi^0$ and $\alpha_{\usig\pi^+}$ for $\ldc\to\usig\pi^+$.
\end{abstract}
\pacs{14.20.Lq, 11.30.Er, 13.88.+e}
\maketitle

\subsection{Introduction}
As hyperon decays, the hadronic weak decay of singly charmed baryons, $\mathcal{B}_c$, is expected to violate parity conservation. In the quark model, $\mathcal{B}_c$ consists of a charm quark ($c$) and two other light quarks ($u$, $d$ and $s$). The ground state $\bc$  has spin $\half$, and decays via the weak interaction, dominantly  producing final states involving a particle with strangeness. For example, the two body decay, $\ldc\to\ld\pi^+$, goes via a $W$-interaction, $c\to W^+ + s$, where $P$-parity is not conserved. $S$- and $P$-waves are allowed between the $\ld$ and $\pi^+$ particles. The parity violation is manifested by the polarization of charmed baryons, which is characterized by the angular distribution of $\ld$ in the $\ldc$ rest frame, taking the form of ${dN\over d\cos\theta_\ld}\propto 1+\alpha_{\ld\pi}\cos\theta_\ld$, where $\alpha_\ldc$ is the decay asymmetry parameter.

Some decay asymmetry parameters in $\bc$ decays, e.g., $\alpha_{\ld\pi}$ for $\ldc\to\ld\pi^+$ and $\alpha_{\Sigma^+\pi^0}$ for $\ldc\to\Sigma^+\pi^0$, have been measured in the FOCUS \cite{focus}, CLEO \cite{cleo1,cleo2,cleo3} and ARGUS~\cite{argus} experiments, as listed in Table~\ref{tab1}. Measurements of $\alpha_{\ld\pi^+}$ are consistent with each other either in the $\ee$ annihilation or photoproduction experiments within one standard deviation. However, their precisions are poor with relative uncertainties larger than 15\%. Average values of these parameters \cite{pdg} are $\alpha_{\ld\pi^+}=-0.91\pm0.15$ and $\alpha_{\Sigma^0\pi^+}=-0.45\pm0.31\pm0.06$. For $\Xi_c^0\to\Xi^-\pi^+$ decays, only one measurement from $\ee$ annihilation was performed, and $\alpha_{\Xi^-\pi^+}$ was measured with relative uncertainty of 41\% as listed in Table~\ref{tab1}.
\begin{table*}[htbp]
\caption{Experimental measurements of decay asymmetry parameters $\alpha_{\ld\pi^+}$ for $\ldc\to\ld\pi^+$,  $\alpha_{\Sigma^+\pi^0}$ for $\ldc\to\Sigma^+\pi^0$ and $\alpha_{\Xi^-\pi^+}$ for
$\Xi^0_c\to\Xi^-\pi^+$. The theoretical predictions are also listed.
\label{tab1}}
\begin{tabular}{lcccc}
\hline\hline
Experiment & Data sets & $\alpha_{\ld\pi^+}$ & $\alpha_{\Sigma^+\pi^0}$ & $\alpha_{\Xi^-\pi^+}$\\\hline
FOCUS\cite{focus} &$\gamma A \sim 80 $GeV&$-0.78\pm0.16\pm0.19$ &&  \\
CLEO \cite{cleo1}  &$\ee\sim\Upsilon(4S)$ & $-0.94^{-0.21+0.12}_{-0.06-0.06}$ & $-0.45\pm0.31\pm0.06$ &  \\
ARGUS\cite{argus}  &$\ee \sim10.4$ GeV &$-0.96\pm0.42$ & & \\
CLEO \cite{cleo2}  &$\ee\sim$10.6 GeV& $-1.1^{+0.4}_{-0.1}$ &  &\\
CLEO \cite{cleo3}  &$\ee\sim\Upsilon(4S)$&        & & $-0.56\pm0.39^{+0.10}_{-0.09}$\\\hline
\multirow{3}{2pt}{Theoretical predictions} &  &$-$0.70\cite{korner}, $-0.67$\cite{xu}  &0.71\cite{korner},0.92\cite{xu} &$-0.38$\cite{korner},$-0.99$\cite{xu}  \\
  & &$-0.95$\cite{cheng},$-0.95$\cite{ivanov}&0.78\cite{cheng},0.43\cite{ivanov} &$-0.38$\cite{cheng},$-0.84$\cite{ivanov}\\
  & &$-0.99$\cite{zenczy},$-0.99$\cite{sharma}&0.39\cite{zenczy},$-0.31$\cite{sharma} &$-0.79$\cite{zenczy},$-0.97$\cite{sharma}\\
\hline\hline

\end{tabular}
\end{table*}

These asymmetry parameters have been predicted by many model calculations, as shown in Table~\ref{tab1}. The calculated decay asymmetry $\alpha_{\ld\pi^+}$ ranges from $-0.67$ to $-0.99$, which is consistent with the current world average $-0.91\pm0.15$ within 2 standard deviations. The agreement between theory and experiment implies the $V$-$A$ structure of the decay process $\ldc\to\ld\pi^+$. The decay asymmetry $\alpha_{\Xi^-\pi^+}$  ranges from $-0.38$ to $-0.99$, which is compatible with the CLEO measurement \cite{cleo3}. However, most model calculations predict the decay asymmetry $\alpha_{\Sigma^+\pi^0}$ to have a positive value, while the measurement from CLEO  \cite{cleo1} gives a negative result of $-0.45\pm0.31\pm0.06$. Additional measurements are important to test the sign of this parameter.

Furthermore, improved measurements with better precisions are desirable to constrain the different model calculations. This is helpful to shed light on the decay mechanism and to test the $CP$ symmetry in charmed baryon sector \cite{focus}.
Based on a data sample with integrated luminosity of 567 pb$^{-1}$ at $\sqrt s=4.6$ GeV accumulated at BESIII recently \cite{datasets}, large statistics of $\ldc$ events are available through the process $\ee\to\ldc\ldcb$. In this paper, we present a proposal to study the decay asymmetry parameters at $\ee$ experiments in the $\tau$-charm energy region, and estimate the sensitivities of measuring $\alpha_{\ld\pi^+},~\alpha_{\Sigma^0\pi^+}$ and $\alpha_{\Xi^-\pi^+}$ parameters, based on the current BESIII data set.

\subsection{Charmed baryon decay asymmetry}
In the process of $\ee\to\gamma^*\to \mathcal{B}_c \bar{\mathcal{B}}_c$,  the charmed baryon $\mathcal{B}_c$ pairs are produced from the electromagnetic process. Hence, they are unpolarized, if we ignore the $Z$-boson contribution, since the $Z$ mass is far from the $\tau$-charm energy region.
A direct measurement on the $\ldc$ polarization was performed at $\Upsilon(4S)/(5S)$ energy points \cite{cleo2}, and the results were consistent with the expectation of $\ldc$ unpolarized production. Based on the unpolarized charmed baryons in $\ee$ annihilation experiment, we will present formulae to measure the $\ldc$ decay asymmetry parameters in two-body hadronic decays, {\it i.e.} $\ldc\to\ld\pi^+,\Sigma^0\pi^+, p\bar K_0$. In addition, we will extend the discussions to other singly charmed baryon decays, such as $\Xi_c\to\Xi^-\pi^+$ and $\Omega_c^0\to\Omega^-\pi^+$.

\subsubsection{$\ldc\to\ld\pi^+$ and $\Sigma^0\pi^+$}

Parity violations in the weak decays of charmed baryons, such as $\ldc\to\ld\pi^+$ and $\Sigma^0\pi^+$, give rise to the polarization of the produced hyperons in the final states. For the unpolarized $\ldc$, the decay asymmetry parameters, $\alpha_{\ld\pi^+}$ and $\alpha_{\Sigma^0\pi^+}$, cannot be observed in the hyperon angular distributions, but they would be related to the hyperon polarizations. This implies that one needs to investigate the hyperon decays to probe the hyperon polarization, and then study the $\ldc$ asymmetry parameters. Experimentally, these hyperon states are reconstructed with the decays $\ld\to p\pi^-,\textrm{~and~}\Sigma^0\to\gamma\ld,\ld\to p\pi^-$. We choose the nonleptonic decay $\ld\to p\pi^-$ as the polarization analyzer.

The amplitude in the weak decay $\ld\to p\pi^-$ is conventionally constructed with $S$- and $P$-waves, and related to the $\ld$ asymmetry parameter as $\alpha_-={2\textrm{Re}(S^*\cdot P)\over |S|^2+|P|^2}$. Instead, we construct the amplitude under the helicity basis. The angular distribution for the decays of $\ld$ into $p\pi^-$ is defined by
\begin{eqnarray}\label{ldang}
f(\theta_p,\phi_p) =\sum_{\lambda_p=\pm1/2} \rho_{M,M'}D^{\half}_{M,\lambda_p}(\phi_p,\theta_p,0)\nonumber\\
\times D^{\half*}_{M',\lambda_p}(\phi_p,\theta_p,0) |H_{\lambda_p}|^2,
\end{eqnarray}
where $\rho$ is the spin density matrix for $\ld$, and $(\theta_p,~\phi_p)$ is the solid angle of the proton in the $\ld$ helicity system, $M(\lambda_p)$ is the helicity of $\ld(p)$, and $H_{\lambda_p}$ is the helicity amplitude, which is related the asymmetry parameter by $|H_{1/2}|^2=(1+\alpha_-)/2$ and $|H_{-1/2}|^2=(1-\alpha_-)/2$, where $\alpha_-$ is the decay asymmetry parameter for $\ld\to p\pi^-$. After integrating over the azimuthal angle $\phi_p$, Eq.~\eqref{ldang} is reduced to
\begin{eqnarray}
f(\theta_p)&=&\textrm{Tr} \rho R,\nonumber\\
 \textrm{~with~}R&=&\left(\begin{array}{cc}
1+\alpha_-\cos\theta_p&0\\
0 & 1-\alpha_-\cos\theta_p
\end{array}\right).
\end{eqnarray}

The spin density matrix, $\rho$, measures the $\ld$ polarization, and it is determined from the $\ld$ production process. For the decay $\ldc\to\ld\pi^+$, the elements of the $\rho$ matrix are defined by
\begin{equation}\label{rhold}
\rho_{\lambda_1,\lambda_2}=\sum_{\lambda=\pm1/2} D^{\half}_{\lambda,\lambda_1}(\phi_\ld,\theta_\ld,0)D^{\half*}_{\lambda,\lambda_2}(\phi_\ld,\theta_\ld,0)A_{\lambda_1}A^*_{\lambda_2},
\end{equation}
where the sum over the $\ldc$ helicity value, $\lambda=\pm 1/2$, takes the same probability due to unpolarization of $\ldc$, $(\theta_\ld,\phi_\ld)$ is the solid angle with reference to the $\ldc$ helicity system, and $A_\lambda$ is the helicity amplitude of this decay, which is related to the asymmetry parameter by $|A_{+\half}|^2=(1+\alpha_{\ld\pi^+})/2$ and $|A_{-\half}|^2=(1-\alpha_{\ld\pi^+})/2$. After integrating over $\phi_\ld$ in Eq.~\eqref{rhold}, one has
\begin{equation}
\rho=\left(\begin{array}{cc}
1+\alpha_{\ld\pi^+} & 0\\
0& 1-\alpha_{\ld\pi^+}
\end{array}\right).
\end{equation}
Then the joint angular distribution for $\ee\to\ldc\ldcb,\ldc\to\ld\pi^+,\ld\to p\pi^-$ is calculated to be \cite{cleo2}
\begin{equation}\label{angldpi}
{dN\over d\cos\theta_p}\propto 1+\alpha_{\ld\pi^+}\alpha_-\cos\theta_p,
\end{equation}
where $\alpha_-$ is the decay asymmetry parameter for $\ld\to p\pi^-$.

It is natural to apply the above formulae to other singly charmed baryon decays like $(\Lambda_c,~\Xi_c,~\Omega_c)(\half^+)\to Y(\half^+)P(0^-)$, $Y\to \mathcal{B}(\half^+)P(0^-)$, where $P$ denotes a pseudoscalar particle, and $Y$ is a hyperon particle, decaying to a baryon ($\mathcal{B}$) and a pseudoscalar particle. The decay asymmetry parameters, $\alpha_{\ld\pi^+}\alpha_-$ in Eq. (\ref{angldpi}), are replaced with those in the $\Lambda_c,~\Xi_c,~\Omega_c$ and subsequent hyperon decays. For example, for the decay $\ldc\to\Sigma^+\pi^0,~\Sigma^+\to p\pi^0$, its angular distribution has a similar form to \cite{cleo1}
\begin{equation}\label{angSigmapi0}
{dN\over d\cos\theta_p}\propto 1+\alpha_{\Sigma^+\pi^0}\alpha_{p\pi^0}\cos\theta_p,
\end{equation}
where $\alpha_{p\pi^0}$ is the decay asymmetry parameter for $\Sigma^+\to p\pi^0$.

For the case of $\ld$ production from $\ldc(\lambda)\to\Sigma^0(\lambda_2)\pi^+,~\Sigma^0\to\ld(\lambda_3)\gamma(\lambda_4)$, where $(\lambda_i)$ represents the helicity value for each particle. The elements of the spin density matrix are defined by
\begin{eqnarray}
\rho_{\lambda_3,\lambda'_3}&=&\sum_{\lambda=\pm1/2} D^{\half}_{\lambda,\lambda_2}(\phi_\usig,\theta_\usig,0)D^{\half*}_{\lambda,\lambda_2}(\phi_\usig,\theta_\usig,0)\nonumber\\
&\times&D^{\half}_{\lambda_2,\lambda_3-\lambda_4}(\phi_\ld,\theta_\ld,0)D^{\half*}_{\lambda'_2,\lambda'_3-\lambda_4}(\phi_\ld,\theta_\ld,0)\nonumber\\
&\times&A_{\lambda_2}A^*_{\lambda'_2}B_{\lambda_3,\lambda_4}B^*_{\lambda'_3,\lambda_4}, \end{eqnarray}
where $(\theta_\usig,\phi_\usig)$ is the $\usig$ solid angles in the $\ldc$ helicity system, whose $z$-axis is taken along the direction of $\usig$ flight, and $A_\lambda$ is the helicity amplitude of the decay $\ldc\to\usig\pi^+$, which is related to the asymmetry parameter by $|A_{+\half}|^2=(1+\alpha_{\usig\pi^+})/2$ and $|A_{-\half}|^2=(1-\alpha_{\usig\pi^+})/2$.
$(\theta_\ld,\phi_\ld)$ is the $\ld$ solid angles in the $\usig$ helicity system, following the convention that the $z$-axis is taken along the direction of $\ld$ flight, and $B_{\lambda_3,\lambda_4}$ is the helicity amplitude of the decay $\usig\to\ld\gamma$. The parity conservation in this radiative decay implies that $B_{\half,1}=B_{-\half,-1}$. After integrating over angles $\phi_\usig$ and $\phi_\ld$, one has
\begin{equation}
\rho=\left(\begin{array}{cc}
1-\alpha_{\ldc}\cos_\ld & 0\\
0 & 1+\alpha_{\ldc}\cos_\ld
\end{array}\right).
\end{equation}
Therefore, the joint angular distribution for the sequential decay $\ee\to\ldc\ldcb,\ldc\to\usig\pi^+,\usig\to\ld\gamma,\ld\to p\pi^-$ reads \cite{cleo1}
\begin{equation}\label{angSigpi}
{dN\over d\cos\theta_\ld d\cos\theta_p}\propto 1-\alpha_{\usig\pi^+}\alpha_{-}\cos\theta_\ld\cos\theta_p.
\end{equation}

\subsubsection{$\xic\to\Xi^-\pi^+,~\Xi^-\to\ld\pi^-$}
For the case of $\ld$ production from $\xic(\lambda)\to\Xi^-(\lambda_2)\pi^+,~\Xi^-\to\ld(\lambda_3)\pi^-$, the elements of the spin density matrix are defined by
\begin{eqnarray}
\rho_{\lambda_3,\lambda'_3}&=&\sum_{\lambda=\pm1/2} D^{\half}_{\lambda,\lambda_2}(\phi_{\Xi^-},\theta_{\Xi^-},0)D^{\half*}_{\lambda,\lambda'_2}(\phi_{\Xi^-},\theta_{\Xi^-},0)\nonumber\\
&\times&D^{\half}_{\lambda_2,\lambda_3}(\phi_\ld,\theta_\ld,0)D^{\half*}_{\lambda'_2,\lambda'_3}(\phi_\ld,\theta_\ld,0)\nonumber\\
&\times&A_{\lambda_2}A^*_{\lambda'_2}B_{\lambda_3}B^*_{\lambda'_3}, \label{eq_Xic}
\end{eqnarray}
where $(\theta_{\Xi^-},\phi_{\Xi^-})$ is the $\Xi^-$ solid angles in the $\xic$ helicity system, whose $z$-axis is taken along the direction of $\Xi^-$ flight, and $A_\lambda$ is the helicity amplitude of the decay $\xic\to\Xi^-\pi^+$, which is related to the asymmetry parameter by $|A_{+\half}|^2=(1+\alpha_{\Xi^-\pi^+})/2$ and $|A_{-\half}|^2=(1-\alpha_{\Xi^-\pi^+})/2$.
$(\theta_\ld,\phi_\ld)$ is the $\ld$ solid angles in the $\Xi^-$ helicity system, where the $z$-axis is taken along the direction of $\ld$ flight, and $B_{\lambda_3}$ is the helicity amplitude of the decay $\Xi^-\to\ld\pi^-$, which is related to the $\Xi^-$ asymmetry parameter by $|B_{+\half}|^2=(1+\alpha_{\Xi^-\pi^+})/2$ and $|A_{-\half}|^2=(1-\alpha_{\Xi^-\pi^+})/2$. Then one has
\begin{widetext}
\begin{equation}
\rho=\left(\begin{array}{cc}
(1+\alpha_{\ld\pi^-})(1+\alpha_{\Xi^-\pi^+}\cos\theta_\ld) & 0\\
0 & (1-\alpha_{\ld\pi^-})(1-\alpha_{\Xi^-\pi^+}\cos\theta_\ld)
\end{array}\right).
\end{equation}
\end{widetext}
The joint angular distribution for the sequential decay $\ee\to\Xi^0_c\bar\Xi_c^0,\Xi_c^0\to\Xi^-\pi^+,\Xi^-\to\ld\pi^-,\ld\to p\pi^-$ reads \cite{cleo1}
\begin{eqnarray}\label{angXipi}
{dN\over d\cos\theta_{\ld} d\cos\theta_p}&\propto& 1+
\alpha_{\Xi^-\pi^+}\alpha_{\ld\pi^-}\cos\theta_{\ld}\nonumber\\
&+&\alpha_-\alpha_{\Xi^-\pi^+}\cos\theta_\ld\cos\theta_p\nonumber\\
&+&\alpha_-\alpha_{\ld\pi^-}\cos\theta_p.
\label{angXipi}
\end{eqnarray}

\subsubsection{$\Omega_c^0\to\Omega^-\pi^+$, $\Omega^-\to\Lambda K^-$}
Let us consider the decay $\Omega_c^0(\half^+)\to\Omega^-({3\over 2}^+)\pi^+$ and $\Omega^-({3\over 2}^+)\to\Lambda(\half^+)K^-(0^-)$. The element of spin density matrix for the $\ld$ is defined by
\begin{eqnarray}
\rho_{\lambda_3,\lambda'_3}&=&\sum_{\lambda,\lambda_2,\lambda'_2} D^{\half}_{\lambda,\lambda_2}(\phi_{\Omega^-},\theta_{\Omega^-},0)D^{\half*}_{\lambda,\lambda'_2}(\phi_{\Omega^-},\theta_{\Omega^-},0)\nonumber\\
&\times&D^{{3\over 2}}_{\lambda_2,\lambda_3}(\phi_\ld,\theta_\ld,0)D^{{3\over 2}*}_{\lambda'_2,\lambda'_3}(\phi_\ld,\theta_\ld,0)\nonumber\\
&\times&A_{\lambda_2}A^*_{\lambda'_2}B_{\lambda_3}B^*_{\lambda'_3},
\end{eqnarray}
where $(\theta_{\Omega^-},\phi_{\Omega^-})$ is the $\Omega^-$ solid angles in the $\Omega_c^0$ helicity system, whose $z$-axis is taken along the direction of $\Omega^-$ flight, and $A_\lambda$ is the helicity amplitude of the decay $\Omega^0_c\to\Omega^-\pi^+$, which is related to the asymmetry parameter by $|A_{+\half}|^2=(1+\alpha_{\Omega^-\pi^+})/2$ and $|A_{-\half}|^2=(1-\alpha_{\Omega^-\pi^+})/2$.
$(\theta_\ld,\phi_\ld)$ is the $\ld$ solid angles in the $\Omega^-$ helicity system under the same convention in Eq.~\eqref{eq_Xic} and $B_{\lambda_3}$ is the helicity amplitude of the decay $\Omega^-\to\ld K^-$, which is related to the $\Omega^-$ asymmetry parameter by $|B_{+\half}|^2=(1+\alpha_{\ld K^-})/2$ and $|A_{-\half}|^2=(1-\alpha_{\ld K^-})/2$.
The joint angular distribution for the sequential decay $\ee\to\Omega_c^0\bar\Omega_c^0,\Omega_c^0\to\Omega^-\pi^+,\Omega^-\to\ld K^-,\ld\to p\pi^-$ reads \cite{cleo1}
\begin{eqnarray}\label{omegac}
{dN\over d\cos\theta_{\Omega^-} d\cos\theta_p}&\propto&
(9\cos^2\theta_{\Omega^-}-5)(\alpha_-\cos\theta_p+\alpha_{\ld K^-})\nonumber\\
&\times&\alpha_{\Omega^-\pi^+}\cos\theta_{\Omega^-}
+(3\cos^2\theta_{\Omega^-}+1)\nonumber\\
&\times&(1+\alpha_-\alpha_{\ld K^-}\cos\theta_p).
\end{eqnarray}
If the angle $\theta_{\Omega^-}$ is integrated out, and we look for the angular distribution of the proton, one has
\begin{equation}
{dN\over d\cos\theta_p}\propto 1+\alpha_-\alpha_{\ld\pi^-}\cos\theta_p.
\end{equation}

\subsubsection{$\ee\to\gamma^*\to\ldc\ldcb$ with $\ldc\to p \bar K^0$}
We consider another situation, where  the strangness in the final state goes to the meson $\bar{K}^0$, e.g., $\ldc\to p \bar{K}^0$. In this process, the proton is produced to be polarized from the $\ldc$ weak decay, but its polarization cannot be analyzed directly, due to having no subsequent decays. In this case, the decay asymmetry parameters could be measured by the correlation of $\ldc$ and $\ldcb$ spins in the process $\ee\to\gamma^*\to\ldc\ldcb$ with $\ldc\to p \bar K^0$ and $\ldcb\to\bar p K^0$. The joint angular distribution is similar to that of process $\ee\to\jp\to\ld\ldb\to(p\pi^-)(\bar p\pi^+)$ \cite{chenh}:
\begin{widetext}
\begin{eqnarray}\label{lld}
&&{dN\over d\cos\theta~d\cos\theta_1d\phi_1
d\cos\bar\theta_1d\bar\phi_1}\nonumber\\&\propto&
4|A_{1/2,1/2}|^2\sin^2\theta[1+\alpha_{p\bar{K}^0}\alpha_{\bar{p}K^0}(\cos\theta_1\cos\bar\theta_1
+\sin\theta_1\sin\bar\theta_1\cos(\phi_1+\bar\phi_1))]\nonumber
\\
&-&2|A_{1/2,-1/2}|^2(1+\cos^2\theta)(\alpha_{p\bar{K}^0}\alpha_{\bar{p}K^0}\cos\theta_1\cos\bar\theta_1-1),
\end{eqnarray}
\end{widetext}
where $A_{1/2,\pm1/2}$ is the helicity amplitude of $\gamma^*\to\ldc\ldcb$, $\theta$ is the polar angle for $\ldc$ in the $\ee$ center-of-mass system, and $\theta_1(\bar\theta_1)$ and $\phi_1(\bar\phi_1)$ are the solid angle of $p(\bar p)$ in the $\ldc(\ldcb)$ helicity system. $\alpha_{p\bar{K}^0}~(\alpha_{\bar{p}K^0})$ is the decay asymmetry parameter for $\ldc\to p\bar{K}^0 (\ldcb\to \bar{p}K^0)$.

\subsubsection{Statistical sensitivity}
With the measured asymmetry parameters of hyperon decays, we can extract the parameters $\alpha_{\ld\pi^+},~\alpha_{\usig\pi^+}$ for $\ldc$ decays, $\alpha_{\Xi^-\pi^+}$ for $\xic$ decays, and parameter product $\alpha_{p\bar{K}^0}\alpha_{\bar{p}K^0}$ by fitting the formulae of joint angular distributions to data. For a given data set with $N$ events observed, a likelihood function is defined by
\begin{equation}
\mathcal{L}=\prod_{i=1}^N\tilde{f}(\theta_i,\alpha),
\end{equation}
where $\tilde{f}$ is a normalized function of angular distribution with helicity angle $\theta_i$ and parameter $\alpha$. The maximum likelihood method is used to estimate the parameter, and its statistical sensitivity is defined by relative statistical uncertainty as
\begin{equation}\label{sensitivity}
\delta(\alpha)={\sqrt{V(\alpha)}\over |\alpha|},
\end{equation}

where $V(\alpha)$ denotes the variance of the parameter $\alpha$. The variance can be determined by
\begin{equation}\label{variance}
V^{-1}(\alpha)=N\int{1\over \tilde{f}(\cos\theta_i,\alpha)}\left[{\partial \tilde{f}(\cos\theta_i,\alpha)\over \partial \alpha}\right]^2 \prod_i \textrm{d}\cos\theta_i,
\end{equation}
where $N$ denotes the observed signal yield.

The parameters, $\alpha_{\ld\pi^+}$ and $\alpha_{\Sigma^+\pi^0}$, are estimated with Eq. (\ref{angldpi}). With the input of the accurate results of $\alpha_-$~\cite{pdg}, their sensitivities are obtained from Eq. (\ref{sensitivity}):

\begin{eqnarray}
{2.13\over \sqrt{N}}\le\delta(\alpha_{\ld\pi^+})\le{3.27\over \sqrt{N}},\nonumber\\
{1.81\over \sqrt{N}}\le\delta(\alpha_{\Sigma^+\pi^0})\le{13.53\over \sqrt{N}}.
\end{eqnarray}

Here we fix the parameters at $\alpha_{\ld\pi^+}=-0.91\pm0.15$ \cite{pdg},  and $\alpha_{\Sigma^+\pi^0}=-0.45\pm0.32$ \cite{pdg}.
The parameter $\alpha_{\usig\pi^+}$ is estimated with Eq. (\ref{angSigpi}), and its sensitivity is calculated by Eq. (\ref{sensitivity})
\begin{equation}
{4.30\over\sqrt{N}} \le \delta(\alpha_{\usig\pi^+})\le{15.01\over\sqrt{N}},
\end{equation}
where the low sensitivity is estimated with the low limit of $\alpha_{\usig\pi^+}=-1$, and the high sensitivity is estimated with  a theoretical prediction $-0.31$ \cite{sharma}.
The parameter $\alpha_{\Xi^-\pi^+}$ is estimated with Eq. (\ref{angXipi}), and its sensitivity is calculated by Eq. (\ref{sensitivity})
\begin{equation}
\delta(\alpha_{\Xi^-\pi^+})={5.38\over\sqrt{ N}}.
\end{equation}
Here all hyperon decay parameters are fixed at PDG values~\cite{pdg}, {\it i.e.}, $\alpha_{\Xi^+\pi^-}=-0.56,~\alpha_{\ld\pi^-}=-0.458$.

The parameter $\alpha_{\Omega^-\pi^+}$ is estimated with Eq. (\ref{omegac}), and its sensitivity is calculated to be
\begin{equation}
\delta(\alpha_{\Omega^-\pi^+})={214.90\over\sqrt{ N}},
\end{equation}
if we take $\alpha_{\Omega^-\pi^+}=0.17$ from a theoretical calculation \cite{omegac}, and $\alpha_-=0.642$, $\alpha_{\Lambda K-}=0.018$ from the PDG \cite{pdg}.

The parameter product $\alpha_{p\bar{K}^0}\alpha_{\bar{p}K^0}$ is estimated with Eq. (\ref{lld}), and according to Eq. (\ref{sensitivity}), the statistical sensitivity reads

\begin{equation}\label{llsen}
\delta (\alpha_{p\bar{K}^0}\alpha_{\bar{p}K^0})\approx{1\over |\alpha_{p\bar{K}^0}\alpha_{\bar{p}K^0}|}\sqrt{{9(d-2)^{2}(d+1)\sqrt{d^2-4}\over
N\left\{48d^2i\tanh^{-1}{\left(\sqrt{2-d\over
2+d}\right)}+(d+2)[(d-9)d+2]\sqrt{d^2-4}\right\}}},
\end{equation}
where $d={2(1+\beta)\over
1-\beta}$, and $\beta$ is the angular distribution parameter for $\ldc$ in the $\ee$ rest frame, which takes the form ${dN\over d\cos\theta}\propto1+\beta\cos^2\theta$ with $\beta={|A_{1/2,-1/2}|^2-2|A_{1/2,1/2}|^2\over |A_{1/2,-1/2}|^2+2|A_{1/2,1/2}|^2 }$. If we take
$|\alpha_{p\bar{K}^0}\alpha_{\bar{p}K^0}|=1$ from a theoretical prediction \cite{apk}, and assume $\beta=-0.3$, we have
\begin{equation}
\delta (\alpha_{p\bar{K}^0}\alpha_{\bar{p}K^0})={8.54 \over \sqrt{N}}.
\end{equation}

To be more straightforward for scaling, we estimate these statistical sensitivities based on a sample of integrated luminosity $\mathcal{L}=1$ fb$^{-1}$, and the number of observed events is calculated with
$$
N=\mathcal{L}\sigma \mathcal{B} \epsilon,
$$
where $\sigma$, $\mathcal{B}$ and $\epsilon$ denote the cross section, combined branching fraction and detection efficiency, respectively. The results, in Table~\ref{mysens}, indicate that one needs a huge data sample to improve the sensitivities for measuring
$\alpha_{\Xi^-\pi^+}$, $\alpha_{\Omega^-\pi^+}$ and $\alpha_{p\bar{K}^0}\alpha_{\bar{p}K^0}$.

If we assume that the $\ldc,~\Xi_c^0$ and $\Omega_c^0$ decays conserve the $CP$ transformation, the charged conjugate events can be combined to measure the asymmetry parameters, so that the sensitivities in Table~\ref{mysens} are improved by a factor of $1/\sqrt 2$.
Based on the 567 pb$^{-1}$ data sample taken at 4.6 GeV collected at BESIII,
an analysis \cite{datasets} has shown that there are $707\pm27$ signal events for $\ldc(\ldcb)\to\ld\pi^+(\bar\ld\pi^-)$, and $309\pm24$ events for $\ldc(\ldcb)\to\Sigma^+\pi^0(\bar\Sigma^-\pi^0)$,
and a precision of (8.0$\sim$12.3)\% for $\alpha_{\ld\pi^+}$, and (10.3$\sim$77.0)\% for $\alpha_{\Sigma^+\pi^0}$ will be achieved. This data sample will provide a first measurement of the asymmetry parameter, $\alpha_{\usig\pi^+}$, with a precision of (18.8$\sim$65.6)\% determined from observed $522\pm27$ events of $\ldc(\ldcb)\to\Sigma^0\pi^+(\bar\Sigma^0\pi^-)$ \cite{datasets}. To resolve the sign issue in the decay $\ldc\to\Sigma^+\pi^0$, the statistical uncertainty of $\alpha_{\Sigma^+\pi^0}$ is required to be at least less than the systematic uncertainty of 13\% \cite{cleo1}, and a data sample is required with a integrated luminosity large than 0.5 fb$^{-1}$.
To study the decays of heavier charmed baryons $\xic$ and $\Omega_c^0$ at BESIII,  one needs to collect data going beyond 4.6 GeV.

\begin{table}
\caption{Estimation of statistical sensitivities $\delta(\alpha)$ with a sample of integrated luminosity of 1 fb$^{-1}$, taken at the energy point $\sqrt s$.\label{mysens}}
\begin{tabular}{lllllll}
\hline\hline
Decay & $\ldc\to\ld\pi^+$ & $\ldc\to\Sigma^+\pi^0$& $\ldc\to\Sigma^0\pi^+$ & $\Xi_c^0\to \Xi^-\pi^+$ & $\Omega_c^0\to \Omega^-\pi^+$ & $\ldc\ldcb\to p\bar{K}^0\bar{p}K^0$ \\\hline
Parameter & $\alpha_{\ld\pi^+}$ &$\alpha_{\Sigma^+\pi^0}$ & $\alpha_{\Sigma^0\pi^+}$ & $\alpha_{\Xi^-\pi^+}$ & $\alpha_{\Omega^-\pi^+}$ & $\alpha_{p\bar{K}^0}\alpha_{\bar{p}K^0}$\\\hline
$\sqrt s $ &4.6 GeV&4.6 GeV&4.6 GeV&10.5 GeV&10.6 GeV&4.6 GeV\\\hline
$\sigma\cdot\mathcal{B}$& 1.88 pb \cite{datasets}& 2.80 pb\cite{datasets}&1.92 pb \cite{datasets} &0.77 pb\cite{argus2} & 11.3 fb\cite{cleoOmegac} & 0.05 pb\cite{datasets}  \\\hline
$\epsilon$&42.2\%\cite{datasets}& 23.8\%\cite{datasets} &29.9\%\cite{datasets}&10.8\%\cite{argus2}&15\%\cite{cleoOmegac} &31.2\%\cite{datasets} \\\hline
$N$       & 793 &666 & 574  & 83 &2 &  16 \\\hline
$\delta(\alpha)$ & $(7.6\sim11.6)$\% &$(7.0\sim52.4)$\%  &$(17.9\sim62.0)$\% & 59\% &152.0\% &213.5\% \\\hline\hline
\end{tabular}
\end{table}

\subsection{Summary}
To summarize, we have presented several approaches to measure the decay asymmetry parameters in the two-body hadronic weak decay of singly charmed baryons, such as $\ldc\to\ld\pi^+,~\usig\pi^+,~\Sigma^0\pi^+$, $p\bar{K}^0$, $\xic\to\Xi^-\pi^+$ and $\Omega_c^0\to\Omega^-\pi^+$. With the reasonable assumption that the charmed baryons produced from $\ee$ annihilations are unpolarized, one can probe the asymmetry parameter by measuring the polarization of the subsequent hyperons via the decay $\ld\to p\pi^-$. We present formulae of joint angular distributions for these processes, which will be used to fit to data to extract the asymmetry parameters.
The sensitivities of the proposed measurements are estimated. Based on the 567 pb$^{-1}$ data collected at BESIII, a rough precision of 10\% for $\alpha_{\ld\pi^+}$ should be achieved, and the sensitivity for measuring $\alpha_{\usig\pi^+}$ for $\ldc\to\usig\pi^+$, is estimated to be (19$\sim$66)\%. A huge data sample is crucial to improve the precision of these parameters in the future and even to study the CP violation of the decay asymmetries of charmed baryons and their anti-baryons.

\vspace{0.5cm} {\bf Acknowledgement:} We are grateful to Prof. Cheng Hai-Yang for helpful communications to check the formula.  This work is partly supported
by the National Natural Science Foundation of China under Contracts No. 11175146, No. 11275266, No. 11235011, No. 11375205, No.11425524, No. 11505010  and No.11565006; Beijing Municipal Government Foundation under Contracts No. KM201610017009 and No. 2015000020124G064; the Youth Innovation Promotion Association of CAS.

\end{CJK*}
\end{document}